\documentclass{amsart}
\usepackage{amsfonts}

\theoremstyle{plain}

\newtheorem{lemma}{Lemma}

\numberwithin{equation}{section}

\begin{document}
\Large
\title[Asymptotic normality of sums of extreme values]{A note on the
asymptotic normality of sums of extreme values}
\author{Gane Samb LO}

\begin{abstract}
Let $X_1$, $X_2$,... be a sequence of independent random variables with
common distribution  function $F$ in the domain of attraction of a Gumbel
extreme value distribution and for each  integer $n\geq 1$, let $X_{1,n}
\leq ... X_{n,n}$ denote the order statistics based on the first $n$ of
these random  variables. Along with related results it is shown that for any
sequence of positive integers $k_n \rightarrow +\infty$ and  $k_{n}/n
\rightarrow 0$ as $n \rightarrow 0$ the sum of the upper $k_n$ extreme
values $X_{n-k_{n},n}+...+X_{n,n}$, when  properly centered and normalized,
converges in distribution to a standard normal random variable $N(0, 1)$. These results constitute an extension of results by S. Cs\"{o}rg\H{o} and
D.M. Mason (1985).\\

\noindent LSTA, Universit\'e Pierre et Marie Curie, Paris, FRANCE\\
\noindent LERSTAD, Universit\'e Gaston Berger, Saint-Louis, SENEGAL\\
\noindent 1178 Evanston Drive NW, T3P 0J9, Calgary, CANADA\\ 
\noindent Emails : gane-samb.lo@ugb.edu.sn, ganesamblo@ganesamblo.net\\

\noindent This paper was published in \textit{Journal of Statistical Planning and Inference}, 22, 1989, 127-136\\
\noindent It was recommended by P. Deheuvels, Institute of France.
\end{abstract}

\maketitle

\section{Introduction}

Let $X_{1},X_{2},....$ be a sequence of independent random variables with common distribution function $F$ and for each integer $n\geq 1$, let $%
X_{1,n}\leq ...\leq X_{n,n}$ denote the order statistics based on the first $%
n$ of these random variables. Cs\"{o}rg\H{o} and Mason (1985, 1986) have recently shown among other results that if

\begin{equation}
1-F(x)=L^{\ast }(x)x^{-a}\text{ }as\text{ }x\longrightarrow \infty, \label{r1}
\end{equation}

\noindent where $L^{*}$ is a slowly varying function at infinity and $a\geq 2$, or if $F$ has exponential-like upper tails, meaning

\begin{equation}
\int_{x}^{+\infty }(1-F(y))dy/(1-F(x))\rightarrow c\text{ }as\text{ }%
x\longrightarrow \infty ,  \label{r2}
\end{equation}

\noindent where $0<c<+\infty$, then for any sequence of integers satisfying

\begin{equation*}
1\leq k_{n}\leq n,\text{ }k_{n}\rightarrow +\infty \text{ }and\text{ }k_{n}/n\rightarrow 0\text{ }as\text{ }n\rightarrow +\infty, \tag{K}
\end{equation*}

\noindent  there exist sequences $A_{n}>0$ of normalizing constants and $C_{n}$ of
centering constants such that%
\begin{equation}
A_{n} \left(\sum_{i=1}^{k_{n}}X_{n-i+1n}-C_{n}\right)\overset{d}{\longrightarrow }N(0,1)%
\text{ }as\text{ }n\rightarrow +\infty. \label{r3}
\end{equation}

\noindent The case \eqref{r1} is contained in the theorem of Cs\"{o}rg\H{o} and Mason (1986) and the case \eqref{r2} is Theorem 1.5 of Cs\"{o}rg\H{o}  and Mason (1985). An application of Theorem 2.4.1 of de Haan (1970) (Lemma \ref{l1} below) combined with Fact 1.4 of Cs\"{o}rg\H{o} and Mason (1985) shows that \eqref{r2} implies the existence of sequences of normalizing constants $a_{n}$ and centering constants $b_{n}$, such that
\begin{equation}
a_{n}^{-1}(X_{n,n}-b_{n})\overset{d}{\longrightarrow }G\text{ }as\text{ }%
n\rightarrow +\infty ,  \label{r4}
\end{equation}

\noindent where $G$ is a Gumbel random variable with distribution function%
\begin{equation*}
P(G\leq x)=\exp (-e^{-x})\text{ for }-\infty <x<+\infty .
\end{equation*}

\noindent Whenever such sequences of constants can be chosen so that \eqref{r4} holds,
we say that $F$ is in the domain of attraction of a Gumbel law, written $%
F\in D(\Lambda)$.\\

\noindent One of the purposes of this note is to show that \eqref{r3} holds more generally than under condition \eqref{r2}, that is, $F\in D(\Lambda)$ is sufficient for \eqref{r3} to hold. This will be a consequence of our main results stated in the next section. We shall also obtain some further extensions of the results of Cs\"{o}rg\H{o} and and Mason. The proofs are given in Section \ref{sec3}.\\

\section{Statement of main results} \label{sec2}

First we introduce some notations. Let

\begin{equation*}
Q(s)=inf\{x:F(x)\geq s),\text{ for }0<s\leq 1,
\end{equation*}

\bigskip \noindent with $Q(0)=Q(0+)$, denote the inverse or quantile function of $F$. Write

\begin{equation*}
\sigma ^{2}=\int_{1-s}^{1}\int_{1-s}^{1}(\min (u,v)-uv)dQ(u)dQ(v),\text{ for 
}0\leq s\leq 1.
\end{equation*}

\noindent For any $0<\beta <\infty $, set

\begin{equation*}
c(s,\beta )=s^{-\beta }\int_{1-s}^{1}(1-u)^{\beta }dQ(u),\text{ for }0\leq
s\leq 1.
\end{equation*}

\noindent For convenience, when $\beta =1$, we set $c(s)=c(s,1)$. (Refer to the next
section for our integral convention).\\

\noindent Let $D^{\ast }(\Lambda )$ denote the subclass of $D(\Lambda )$ consisting of
all distribution functions $F$ whose quantile function $Q$ satisfies%
\begin{equation*}
Q(1-s)=a+\int_{s}^{1}u^{-\beta }r(u)du,
\end{equation*}

\noindent for all $s\geq 0$ sufficiently small, $a$ is a fixed constant and $r$ is a
strictly positive function slowly varying at zero. The fact that $D^{\ast
}(\Lambda )$ is a subclass of $D(\Lambda )$ follows from Theorem 2.4.1 of de
Haan (1970).\\

\noindent For any sequence of positive integers $k_{n}$, such that $(K)$ holds and $%
F\in D(\Lambda )$, set for $n=1,2,...,$

\begin{equation*}
\mu _{n}(k_{n})=\int_{1-k_{n}/n}^{1}Q(s)ds.
\end{equation*}

\noindent The following theorem contains our main results.\\

\noindent \textbf{Theorem}.
\textit{On a rich enough probability space, there exist a sequence of independent
random variables $X_{1},X_{2},....,$ with common distribution function $F$
and a sequence of Brownian bridges $B_{1},$ $B_{2},$, . . . , such that for
any sequence $k_{n}$ satisfying $(K)$, whenever $F\in D(\Lambda )$},

\begin{eqnarray}
&&k_{n}^{1/2}c(k_{n}/n)^{-1}\left\{ \sum_{i=1}^{k_{n}}X_{n-i+1n}-\mu
_{n}(k_{n})\right\}   \label{r5} \\
&=&-(n/k_{n})^{1/2}c(k_{n}/n)^{-1}%
\int_{1-k_{n}/n}^{1}B_{n}(s)dQ(s)+o_{P}(1):=Z_{n}+o_{P}(1),  \notag
\end{eqnarray}%
\textit{and whenever} $F\in D^{\ast }(\Lambda ),$%
\begin{eqnarray}
&&k_{n}^{-1/2}c(k_{n}/n)^{-1}\left\{
\sum_{i=1}^{k_{n}}X_{n-i+1n}-Q(1-k_{n}/n)\right\}   \label{r6} \\
&=&-(n/k_{n})^{1/2}c(k_{n}/n)^{-1}B_{n}(1-k_{n}/n)+o_{P}(1):=Y_{n}+o_{P}(1),
\notag
\end{eqnarray}

\noindent \textit{and}
$$
k_{n}^{-1/2}c(k_{n}/n)^{-1}\left\{
\sum_{i=1}^{k_{n}}X_{n-i+1n}-k_{n}X_{n-k_{n},n}-n%
\int_{1-k_{n}/n}^{1}r(1-s)ds)\right\}
$$
\begin{equation}
=Z_{n}-Y_{n}+o_{P}(1).  \label{r7}
\end{equation}

\noindent\textit{ Furthermore, the random variables on the left side of \eqref{r5}, \eqref{r6}
and \eqref{r7}, respectively, converge in distribution to $N(0,2)$, $N(0,1)$
and $N(0,1)$, respectively, as $n\rightarrow +\infty$.}\\

\noindent \textbf{Remark}. With the choice $A_{n}=\left( 2k_{n}\right) ^{-1/2}c(k_{n}/n)^{-1}$ and $C_{n}=\mu _{n}(k_{n}),$ we see that our theorem implies \eqref{r3} whenever $D(\Lambda)$. Our theorem also extends Theorem 1.5, 1.7 and 2.1 and Corollary 2.5 of Cs\"{o}rg\H{o} and Mason (1985). The random variable on the left side of \eqref{r7} is related to the Hill (1975) estimator of the tail index of a distribution for this random variable was motivated by the work of Mason (1982) (see also Deheuvels, Haeusler and Mason (1988)).

\section{Proof of the theorem} \label{sec3}

We use the following integral convention : When $0\leq a\leq
b\leq 1,$ g is left-continuous and f is right-continuous,

\begin{equation*}
\int_{b}^{a}fdg=\int_{(a,b)}fdg\text{ and }\int_{a}^{b}gdf=\int_{(a,b)}gdf
\end{equation*}

\noindent whenever these integrals make sense as Lebesgue-Stieltjes integrals. In this
case, the usual integration by parts formula

\bigskip 
\begin{equation*}
\int_{a}^{b}fdg+\int_{a}^{b}gdf=g(b)f(b)-g(a)f(a)
\end{equation*}

\noindent is valid.\\

\noindent The proof of our theorem will follow closely the proofs of the results of
Cs\"{o}rg\H{o} and Mason (1985), substituting their technical lemmas concerning
properties of the quantile functions of distribution functions satisfying \eqref{r2}, by those describing properties of the quantile functions of $F\in
D(\Lambda ).$ We therefore begin with these technical lemmas.\\

\bigskip

\begin{lemma} \label{l1} $F\in D(\Lambda )$ if and only if for each choice of $0\leq x,y,w,z<\infty $
\ fixed, $y\neq w,$%
\begin{equation}
\frac{Q\left( 1-sx\right) -Q\left( 1-sz\right) }{Q\left( 1-sy\right)
-Q\left( 1-sw\right) }\rightarrow \frac{\log x-\log z}{\log y-\log w}\text{
\ as s}\downarrow 0.  \label{r8}
\end{equation}
\end{lemma}

\noindent This is Theorem 2.4.1 of de Haan (1970).\\

\begin{lemma} \label{l2} Whenever $F\in D(\Lambda ),$ $c(s,\beta )$ is slowly varying at
zero for each choice of $0<\beta <\infty $ .
\end{lemma}

\noindent  \textbf{Proof}. We have to show that for each $0<\lambda <\infty $ and $%
0<\beta <\infty ,$%
\begin{equation}
c(\lambda s,\beta )/c(s,\beta )\rightarrow 1\ \text{as \ }s\downarrow 0.
\label{r9}
\end{equation}

\noindent Choose any $0<\lambda <\infty $ \ and $0<\theta <1.$ Then for all $s>0$ small enough we have%
\begin{eqnarray}
\int_{1-\lambda s}^{1}\left( 1-u\right) ^{\beta }dQ\left( u\right)
&=&\sum\limits_{i=0}^{\infty }\int_{1-\lambda s\theta ^{i}}^{1-\lambda
s\theta ^{i+1}}\left( 1-u\right) ^{\beta }dQ\left( u\right) \notag \\
&\leq &\sum\limits_{i=0}^{\infty }\lambda ^{\beta }s^{\beta }\theta
^{i\beta }\left\{ Q\left( 1-\lambda s\theta ^{i+1}\right) -Q\left( 1-\lambda
s\theta ^{i}\right) \right\}  \label{r10}
\end{eqnarray}

\noindent Applying Lemma \ref{l1} gives

\begin{equation}
\frac{Q\left( 1-\lambda \theta u\right) -Q\left( 1-\lambda u\right) }{%
Q\left( 1-\theta u\right) -Q\left( 1-u\right) }\rightarrow 1\text{ as s}%
\downarrow 0.  \label{r11}
\end{equation}

\noindent Select any \ $0<\varepsilon <\infty$. From \eqref{r11} we have that, for all $s>0$ sufficiently small, expression \eqref{r10} is

\begin{eqnarray*}
&\leq &\left( 1+\varepsilon \right) \sum\limits_{i=0}^{\infty }\lambda
^{\beta }s^{\beta }\theta ^{i\beta }\left\{ Q\left( 1-\lambda s\theta
^{i+1}\right) -Q\left( 1-\lambda s\theta ^{i}\right) \right\}  \\
&\leq &\frac{\left( 1+\varepsilon \right) \lambda ^{\beta }}{\theta ^{\beta }%
}\sum\limits_{i=0}^{\infty }\theta ^{\left( i+1\right) \beta
}\int_{1-s\theta ^{i}}^{1-s\theta ^{i+1}}dQ\left( u\right)  \\
&\leq &\frac{\left( 1+\varepsilon \right) \lambda ^{\beta }}{\theta ^{\beta }%
}\int_{1-s}^{1}\left( 1-u\right) dQ\left( u\right). 
\end{eqnarray*}

\noindent Thus for all $s>0$ sufficiently small,

\begin{equation}
c(\lambda s,\beta )\leq \frac{\left( 1+\varepsilon \right) }{\theta ^{\beta }%
}c(s,\beta ).  \label{r12}
\end{equation}

\noindent Observing that for all $s>0$ small enough,%
\begin{equation*}
\int_{1-\lambda s}^{1}\left( 1-u\right) ^{\beta }dQ\left( u\right) \geq
\sum\limits_{i=0}^{\infty }\lambda ^{\beta }s^{\beta }\theta ^{\left(
i+1\right) \beta }\left\{ Q\left( 1-\lambda s\theta ^{i+1}\right) -Q\left(
1-\lambda s\theta ^{i}\right) \right\} ,
\end{equation*}

\noindent we see that by an argument very much like the one just given, we have for
all $s>0$ sufficiently small,

\begin{equation}
c(\lambda s,\beta )\geq \left( 1+\epsilon \right) \theta ^{\beta }c(s,\beta).  \label{r13}
\end{equation}

\noindent Assertion \eqref{r9} now follows from inequalities \eqref{r12} and \eqref{r13}
by the fact that $\theta $ can be chosen arbitrarily close to one and $\varepsilon$ arbitrarily close to zero. This completes the proof of Lemma \ref{l2}.\\ 

\noindent The following lemma is related to Theorem 1.4.3.d of de Haan (1970) and its proof is based on a modification of the techniques used to prove this theorem. For details see Deheuvels \textit{et al.} (1986).\\

\begin{lemma} \label{l3} Whenever $F\in D(\Lambda )$, there exists a constant \ $-\infty <b<+\infty $
\ such that for all \ $0<s\leq 1/2,$

\begin{equation}
Q\left( 1-s\right) =b-c(s)+\int_{s}^{1}u^{-1}c(u)du.  \label{r14}
\end{equation}
\end{lemma}

\bigskip

\begin{lemma} \label{l4}
Whenever $F\in D(\Lambda )$, for each \ $0<x<+\infty ,$

\begin{equation}
\lim_{s\downarrow 0}\frac{Q\left( 1-xs\right) -Q\left( 1-s\right) }{c(s)}%
=-\log x.  \label{r15}
\end{equation}
\end{lemma}

\bigskip \noindent \textbf{Proof}. Applying Lemma \ref{l3}, we have for any $0<x<+\infty$ and for all 
$s$ sufficiently small%
\begin{equation*}
\frac{Q\left( 1-xs\right) -Q\left( 1-s\right) }{c(s)}=\frac{c(s)-c(xs)}{c(s)}%
+\int_{xs}^{s}\frac{c(u)}{u}du.
\end{equation*}

\noindent Since $c$ is slowly varying at zero, both

\bigskip 
\begin{equation*}
\inf \left\{ \frac{c(u)}{c(s)}:u\in I(s)\right\} \rightarrow 1
\end{equation*}

\noindent and
\begin{equation*}
\sup \left\{ \frac{c(u)}{c(s)}:u\in I(s)\right\} \rightarrow 1
\end{equation*}

\noindent as s$\downarrow 0$ \ where $I(s)$ is the closed interval formed by $xs$ and $%
s$. From these two facts the proof of Lemma \ref{l4} follows immediately.

\begin{lemma} \label{l5} Whenever $F\in D(\Lambda )$, for each \ $0<\beta <+\infty $%
\begin{equation}
c(s,\beta )/c(s)\rightarrow 1/\beta \text{ as s}\downarrow 0.  \label{r16}
\end{equation}
\end{lemma}

\bigskip \noindent \textbf{Proof}. Let $\widehat{Q}(1-s)=$ $Q(1-s^{1/\beta })$. Since by Lemma \ref{l1} for any
choice of $0<x,y<\infty ,$ $y\neq 1,$%
\begin{equation*}
\frac{\widehat{Q}(1-xu)-\widehat{Q}(1-u)}{\widehat{Q}(1-yu)-\widehat{Q}(1-u)}=\frac{{Q}(1-x^{1/\beta }u^{1/\beta
})-Q(1-u^{1/\beta })}{Q(1-y^{1/\beta }u^{1/\beta })-Q(1-u^{1/\beta })}
\end{equation*}

\noindent converges as $u\downarrow 0$ to $\log x/\log y$, we conclude that $\widehat{Q}\in
D(\Lambda)$. Let  
\begin{equation*}
\widehat{c}(s)=s^{-1}\int_{1-s}^{1}\left( 1-u\right) d\widehat{Q},\text{ \ \ \ for }0<s<1.
\end{equation*}

\noindent A change of variables shows that $\widehat{c}(s^{\beta})=c(s,\beta )$ \ for $0<s<1$. Thus

\begin{eqnarray*}
\frac{c(s,\beta )}{c(s)} &=&\frac{\widehat{c}(s^{\beta })}{c(s)}
\\
&=&\frac{Q(1-2s)-Q(1-s)}{c(s)}\times \frac{\widehat{Q}(1-\left(
2s\right) ^{\beta })-\widehat{Q}(1-s^{\beta })}{Q(1-2s)-Q(1-s)}
\\
&&\times \frac{\widehat{c}(s^{\beta })}{\widehat{Q}(1-\left( 2s\right) ^{\beta })-\widehat{Q}(1-s^{\beta })},
\end{eqnarray*}

\noindent which by Lemmas \ref{l1} and \ref{l4} converges to $1/\beta $ as s$\downarrow 0$,
completing the proof of Lemma \ref{l5}.

\bigskip

\begin{lemma} \label{l6}
Whenever $F\in D(\Lambda )$,%
\begin{equation}
\sigma ^{2}(s)/\left( 2sc^{2}(s)\right) \rightarrow 1\text{, \ \ as }%
s\downarrow 0.  \label{r17}
\end{equation}
\end{lemma}

\bigskip \noindent \textbf{Proof. }The proof is based on Lemma \ref{l5} and follows almost exactly as
the proof of Lemma 3.3 of Cs\"{o}rg\H{o} and Mason (1985). Therefore, the details
are omitted. The proof of the following lemma is an easy consequence of the
Karamata representation for a slowly varying function.

\begin{lemma} \label{l7}
Let $a_{n}$ be any sequence of positive constants such that $%
a_{n}\rightarrow 0$ and \ $na_{n}\rightarrow \infty .$ Also let $L$ be any
slowly varying function at zero. Then for any $0<\beta <\infty ,$%
\begin{equation}
n^{-\beta }L\left( 1/n\right) /\left( \left( a_{n}\right) ^{\beta }L\left(
a_{n}\right) \right) \rightarrow 0\text{ \ \ as }n\downarrow \infty. 
\label{r18}
\end{equation}
\end{lemma}

\bigskip \noindent We now describe the probability space on which the assertions of the theorem
are assumed to hold. M. Cs\"{o}rg\H{o}, S. Cs\"{o}rg\H{o} , Horv\'{a}th and Mason (1986) have
constructed a probability space $(\Omega ,\mathcal{A},\mathbb{P})$ carrying
a sequence $U_{1},U_{2},...$ of independent random variables uniformly
distributed on $(0,1)$ and a sequence $B_{1},B_{2},...$ of Brownian
bridges such that for the empirical process 
\begin{equation*}
\alpha _{n}(s)=n^{1/2}\left\{ G_{n}\left( s\right) -s\right\} ,\text{ }0\leq
s\leq 1,
\end{equation*}

\noindent  and the quantile process 
\begin{equation*}
\beta _{n}(s)=n^{1/2}\{s-U_{n}(s)\},0\leq s\leq 1,
\end{equation*}%
where 
\begin{equation*}
G_{n}(s)=n^{-1}\{k:1\leq k\leq n,U_{k}\leq s\},
\end{equation*}

\noindent and, with $U_{1,n}\leq \ldots \leq U_{n,n}$ denoting the order statistics
corresponding to $U_{1},\ldots ,U_{n}.$, 
\begin{equation*}
U_{n}(s)=\left\{ 
\begin{tabular}{lll}
$U_{k,n}$ & $if$ & $(k-1)/n<s\leq k/n,k=1,\ldots ,n,$ \\ 
$U_{1,n}$ & $if$ & $s=0$%
\end{tabular}%
\right. 
\end{equation*}

\noindent we have 
\begin{equation}
\sup_{0\leq s\leq 1}n^{\nu _{1}}|\alpha _{n}(s)-\bar{B}_{n}(s)|/(1-s)^{-\nu
_{1}+1/2}=O_{p}(1)  \label{r19}
\end{equation}

\noindent with $\bar{B}_{n}(s)=B_{n}(s)$ for $1/n\leq s\leq 1-1/n$ and zero elsewhere and 
\begin{equation}  \label{r20}
\sup_{0\leq s \leq 1-1/n }n^{\nu_2}|\beta_n(s)-B_n(s)|/(1-s)^{-\nu_2 +1/2}
=O_p(1)
\end{equation}

\noindent where $\nu_1$, and $\nu_2$ are any fixed number such that $0\leq \nu_1<1/4$
and $0 \leq \nu_2\leq 1/2$. The statement in \eqref{r19} follows from Theorem
2.1, while the statement in \eqref{r20} is easily inferred from Corollaries
2.1 and 4.2.2 of the above paper.\\

\noindent Thoughout the remainder of the proof of our theorem, we assume that we are
on the probability space of Cs\"{o}rg\H{o} et al.(1986). Since the sequence of
random variables $X_1, X_2,\ldots,$ is equal in distribution to $Q(U_1),
Q(U_2,),\ldots$, we can and do assume that the first sequence is equal to
the second.\\

\noindent First assume $F\in D(\Lambda )$. We shall establish \eqref{r5}. Applying
integration by parts we see that the left side of \eqref{r5} equals

$$
-(n/k_n)^{1/2}c(k_n/n)^{-1}\int_{1-k_n/n}^1\alpha_n(s)dQ(s)
$$
$$
+nk_n^{-1/2}%
\int_{U_n-k_n/n}^{1-k_n/n}(1-G_n(s)-k_n/n )\frac{dQ(s)}{c(k_n/n)}
$$
$$
:=\Delta_{1,n}+\Delta_{2,n}.
$$

\noindent We shall first show that 
\begin{equation*}
\Delta_{1,n}=Z_n+R_n
\end{equation*}

\noindent with $R_n=o_p(1)$. From \eqref{r19}  we have for any $0<\nu<1/4$, 
\begin{equation}  \label{r21}
\sup_{ 0\leq s\leq 1 }n^{\nu_1}|\alpha_n(s)-\bar{B}_n(s)|/(1-s)^{-\nu_1
+1/2}=O_p(n^{-\nu}).
\end{equation}

\noindent Notice that for any such $\nu$,

\begin{eqnarray*}
&& |R_n|\leq \sup_{ 0\leq s\leq 1 } \frac{|\alpha_n(s)-\bar{B}_n(s)|}{%
(1-s)^{-\nu_1 +1/2}}\left( \frac{n}{k_n}\right)^{1/2}\int_{1-k_n/n}^1(
1-s)^{-\nu_1 +1/2}dQ(s)/c(k_n/n) \\
&&+ \left\vert \left( \frac{n}{k_n}\right)^{1/2}\int_{1-1/n}^1
B_n(s)dQ(s)/c(k_n/n)\right \vert :=R_{1,n}+R_{2,n}.
\end{eqnarray*}

\noindent From \eqref{r21}, we obtain 
\begin{equation*}
R_{1,n}=O_p(n^{-\nu}n^\nu \frac{c(k_n/n,1/2-\nu)}{c(k_n/n)}k_n^{-\nu},
\end{equation*}

\noindent which by Lemma \ref{l5} equals $o_p(1)$. Also

\begin{equation*}
E R_{2,n}^2=\sigma^2(1/n)/\left(\frac{k_n}{n}c_n^2\left(\frac{k_n}{n}\right)
\right)
\end{equation*}

\noindent which by Lemma \ref{l6} is 
\begin{equation*}
\sigma ^{2}(1/n)/\sigma ^{2}(k_{n}/n)\ \ as\ n\rightarrow \infty. 
\end{equation*}

\noindent From Lemmas \ref{l2} and \ref{l6} we infer that $\sigma^2(s)$ is regularly varying of
exponent one at zero. Hence, by Lemma \ref{l7},

\begin{equation*}
\sigma ^{2}(1/n)/\sigma ^{2}(k_{n}/n)\rightarrow 0\ \ as\ n\rightarrow
\infty 
\end{equation*}

\bigskip \noindent which yields $R_{2,n}=o_p(1)$. Thus we have proved $R_{n}=o_p(1)$.\newline
Next we show that $\Delta_{2,n}=o_p(1)$. Choose any $1 <\lambda<\infty$ and
set

\begin{equation*}
T_{n}(\lambda
)=nk_{n}^{-1/2}c(k_{n}/n)^{-1}|1-G_{n}(1-k_{n}/n)-k_{n}/n|\{Q(r_{n}^{+}(%
\lambda ))-Q(r_{n}^{-}(\lambda ))\},
\end{equation*}%
\begin{equation*}
r_{n}^{-}(\lambda ))=1-\frac{\lambda k_{n}}{n}\ \ and\ r_{n}^{+}(\lambda
))=1-\frac{k_{n}}{\lambda n}.
\end{equation*}

\noindent Notice that since for all $s$ in the closed interval formed by $%
U_{n-k_{n},n}$ and $1-k_{n}/n$,

\begin{equation*}
|1-G_{n}(s)-k_{n}/n|\leq |1-G_{n}(1-k_{n}/n)-k_{n}/n|
\end{equation*}

\noindent we have for any $1<\lambda <\infty $ 
\begin{equation*}
\liminf_{n\rightarrow \infty }P(|\Delta _{2,n}|\leq T_{n}(\lambda )|\geq
\liminf_{n\rightarrow \infty }P(r_{n}^{-}(\lambda )\leq U_{n-k_{n},n}\leq
r_{n}^{+}(\lambda )).
\end{equation*}

\noindent Since $(K)$ implies (cf. Balkema and de Haan (1975)) that 
\begin{equation}
n(l-U_{n-k_{n},n)/k_{n}}\rightarrow ^{P}1\text{ as } n\rightarrow \infty: 
\label{r22}
\end{equation}

\noindent the lower bound in the above inequality equals one. Hence for each $%
1<\lambda <\infty $,

\begin{equation}
\lim_{n\rightarrow }P(|\Delta _{2,n}|\leq |T_{n}(\lambda )|)=1.  \label{r23}
\end{equation}

\noindent Observe for each $1<\lambda <\infty $ 
\begin{equation*}
E(T_{n}(\lambda )\leq \left\{ Q\left( 1-\frac{k_{n}}{\lambda _{n}}\right)
-Q\left( 1-\frac{\lambda k_{n}}{n}\right) \right\} /c(k_{n}/n).
\end{equation*}

\noindent Applying Lemma \ref{l4} we see that this last expression converges to $2\log
\lambda $, which yields

\begin{equation}
\lim_{\lambda \downarrow 1}\limsup_{n\rightarrow \infty }E(T_{n}(\lambda )=0.
\label{r24}
\end{equation}

\noindent The fact that $\Delta _{2,n}=o_{p}(1)$ now follows by an elementary argument
based on \eqref{r23} and \eqref{r24}. This completes the proof of \eqref{r5}.\\

\noindent Next consider \eqref{r6}. Notice that since $F\in D^{\star }(\Lambda )$,

\begin{equation*}
c(s)=s^{-1}\int_{1-s}^1r(1-u)du.
\end{equation*}

\noindent Thus since $r>0$ and slowly varying at zero Theorem 1.2.1 of de Haan (1970)
gives 
\begin{equation}  \label{r25}
r(s)/c(s) \rightarrow 1 \ \ \mbox{as}\ \ s\downarrow 0
\end{equation}

\noindent The left side of \eqref{r6} equals 
\begin{eqnarray*}
&& -\frac{k_n^{1/2}}{c(k_n/n}\int_{k_n/n}^{1-U_{n-k_n/n}}\frac{r(u)}{u}%
du=-k_n^{1/2}\frac{r(k_n/n)}{c(k_n/n)}\{\log(1-U_{n-k_n/n}\} \\
&& \ \ \ \ - \ \ \frac{k_n^{1/2}}{c(k_n/n}%
\int_{k_n/n}^{1-U_{n-k_n/n}}(r(u)-r(k_n/n))\frac{du}{u}:=\Delta_{1,n}^\star
+\Delta_{2,n}^\star
\end{eqnarray*}

\noindent The same argument based on \eqref{r20} as given in Cs\"{o}rg\H{o} and Mason (1985)
shows that 
\begin{equation*}
-k_n^{1/2}(\log(1-U_{n-k_n/n})-\log(k_n/n))=Y_n+o_p(1).
\end{equation*}

\noindent Therefore by \eqref{r25} and the fact that $Y_n=o_p(1)$  we have  
\begin{equation*}
\Delta_{1,n}^\star =Y_n +o_p(1).
\end{equation*}

\noindent Since $r$ is slowly varying at zero, we get for each $1<\lambda <\infty$ as $%
n \rightarrow \infty$,

\begin{equation}
\sup\left\{|r(s)-r(k_n/n)/c(k_n/n):\frac{k_n}{\lambda_n}\leq s \leq \frac{%
\lambda k_n}{n} | \right\} \rightarrow 0  \label{r26}
\end{equation}

\noindent The fact that $\Delta_{2,n}^\star=o_p(1)$ now follows easily from $Y_n=o_p(1)
$, \eqref{r22},\eqref{r26} completing the proof of \eqref{r6}.\\

\noindent Since $F\in D^\star(\Gamma)$ we have

\begin{equation*}
\mu_n(k_n)-k_nQ(1-k_n/n)=\int_{1-k_n/n}^1r(1-s)ds.
\end{equation*}

\noindent Assertion (7) is now a direct consequence of \eqref{r5} and \eqref{r6}.\\

\noindent Finally we prove the convergence in distribution of $Z_n$, $Y_n$, and $Z_n-
Y_n$,, to $N(0,2), N(0,1)$ and $N(0,1)$, respectively, as $n \rightarrow
\infty$. Notice that the $Z_n$, random variable in \eqref{r5} is normal with
mean zero and second moment $\sigma^2(k_n/n)/(k_nc^2(k_n)/n),$, which by
Lemma \ref{l6} converges to $1$ as $n\rightarrow \infty$. The $Y_n$, random
variable in \eqref{r6} is normal with mean zero and second moment $1-k_n/n
\rightarrow 1$ as $n\rightarrow \infty$.\\

\noindent The $Z_{n}-Y_{n}$ random variable in \eqref{r7} is normal with mean zero.
Applying Lemmas \ref{l5} and \ref{l6} it is easy to verify that $E(Z_{n}-Y_{n})^{2}%
\rightarrow 1$ and $n\rightarrow \infty $. This completes the proof of the
theorem.

\bigskip \noindent{\textbf{Acknowledgement}}.\newline

\noindent The author is extremely grateful to David Mason, Erich Haeusler
and Paul Deheuvels, for helping him to complete this work.

\end{document}